\documentclass[aps,prl,floatfix,twocolumn,superscriptaddress,showpacs]{revtex4}  
\usepackage{graphicx}  
\usepackage{dcolumn}   
\usepackage{bm}        
\usepackage{amssymb,amsmath}   
\usepackage[utf8x]{inputenc}

\begin{document}

\title{Radiative heat transfer between two dielectric nanogratings in the scattering approach}


\author{J. Lussange}
\affiliation{Laboratoire Kastler-Brossel, CNRS, ENS, UPMC, Case 74, F-75252 Paris, France}
\author{R. Gu\'erout}
\affiliation{Laboratoire Kastler-Brossel, CNRS, ENS, UPMC, Case 74, F-75252 Paris, France}
\author{F.S.S. Rosa}
\affiliation{Laboratoire Charles Fabry, Institut d’Optique, CNRS, Universit\'e Paris-Sud, Campus Polytechnique, RD128, F-91127 Palaiseau Cedex, France}
\author{J.-J. Greffet}
\affiliation{Laboratoire Charles Fabry, Institut d’Optique, CNRS, Universit\'e Paris-Sud, Campus Polytechnique, RD128, F-91127 Palaiseau Cedex, France}
\author{A. Lambrecht}
\affiliation{Laboratoire Kastler-Brossel, CNRS, ENS, UPMC, Case 74, F-75252 Paris, France}
\author{S. Reynaud}
\affiliation{Laboratoire Kastler-Brossel, CNRS, ENS, UPMC, Case 74, F-75252 Paris, France}


\begin{abstract}
We present a theoretical study of radiative heat transfer between dielectric 
nanogratings in the scattering approach. As a comparision with these exact results, 
we also evaluate the domain of validity of Derjaguin's Proximity Approximation (PA). 
We consider a system of two corrugated silica plates with
various grating geometries, separation distances, and lateral
displacement of the plates with respect to one another. Numerical
computations show that while the PA is a good approximation for
aligned gratings, it cannot be used when the gratings are laterally
displaced. We illustrate this by a thermal modulator device for
nanosystems based on such a displacement. 
\end{abstract}

\maketitle

Recent experiments and theoretical work have given promising perspectives in the field of radiative heat transfer
in the micrometer range~\cite{Biehs2010,Otey2011,Svetovoy2012}. It has been shown that radiative heat transfer greatly exceeds the black body limit
for distances shorter than the average thermal wavelength, which is understood as an effect arising from the contribution of the evanescent waves.
The studies of near-field heat transfer are of great interest to the design of both NEMS and MEMS which are naturally affected by possible
side-effects of heat exchange at the nanoscale. Other potential applications lie in the fields of nanotechnology, photonic
crystals~\cite{BenAbdallah2010}, metamaterials~\cite{Joulain2010,Francoeur2011}, thermalphotovoltaics~\cite{Park2008,Laroche2006}, multilayered 
structures~\cite{Biehs2007}, improved resolution in nano-structure imaging, and new nano-fabrication techniques.

While radiative heat transfer beyond Stefan-Boltzmann's law was
observed experimentally~\cite{Hargreaves1969} and described
theoretically~\cite{Polder1971} over forty years ago, radiative heat
transfer between two parallel flat plates at the nanoscale has been
considered experimentally only
recently~\cite{Volokitin2008,Rousseau2009,Shen2009}. The most
interesting features of this field are the possible side-effects of
non-trivial geometries on the thermal emission of nano-objects. Thus
an in-depth study of heat transfer for different configurations has
been performed over the years, ranging from the case of a particle
facing a surface~\cite{Buhmann2008,Huth2010,Antezza2008,Mulet2001},
to particles or nanospheres facing each
other~\cite{Narayanaswamy2008,Sasihithlu2011,Domingues2005,Sherkunov2009,Chapuis2008},
or more recently to the sphere-plane
geometry~\cite{Otey2011,Emig2011}. One should also note that for
nearly flat surfaces where roughness is considered as a perturbation
factor, certain perturbative approaches can be
used~\cite{Derjaguin1934,Blocki1977,Biehs2011}. But for larger geometrical
irregularities, more accurate methods become
necessary~\cite{McCauley2012}. These more complex geometries are
best described through a scattering
approach~\cite{Bimonte2009,Messina2011,Emig2011,Guerout2012}.
Another exciting perspective is the study of the variation in heat
transfer brought forth by surface polaritons in certain
materials~\cite{Mulet2001}. In this paper we focus on the interplay
between the surface waves excitation and the surface profile, as
shown in Fig.1.

The fact that the radiative heat transfer in near-field considerably
changes with variation of the separation distance between plane
surfaces has already been
shown~\cite{Biehs2010,Otey2011,VanZwol2011}. When introducing a
profile for the interfaces, the flux is expected to depend on the
relative lateral displacement of the two surfaces denoted $\delta$,
as seen in Fig.1.
\begin{figure}[h]
\includegraphics[scale=0.41]{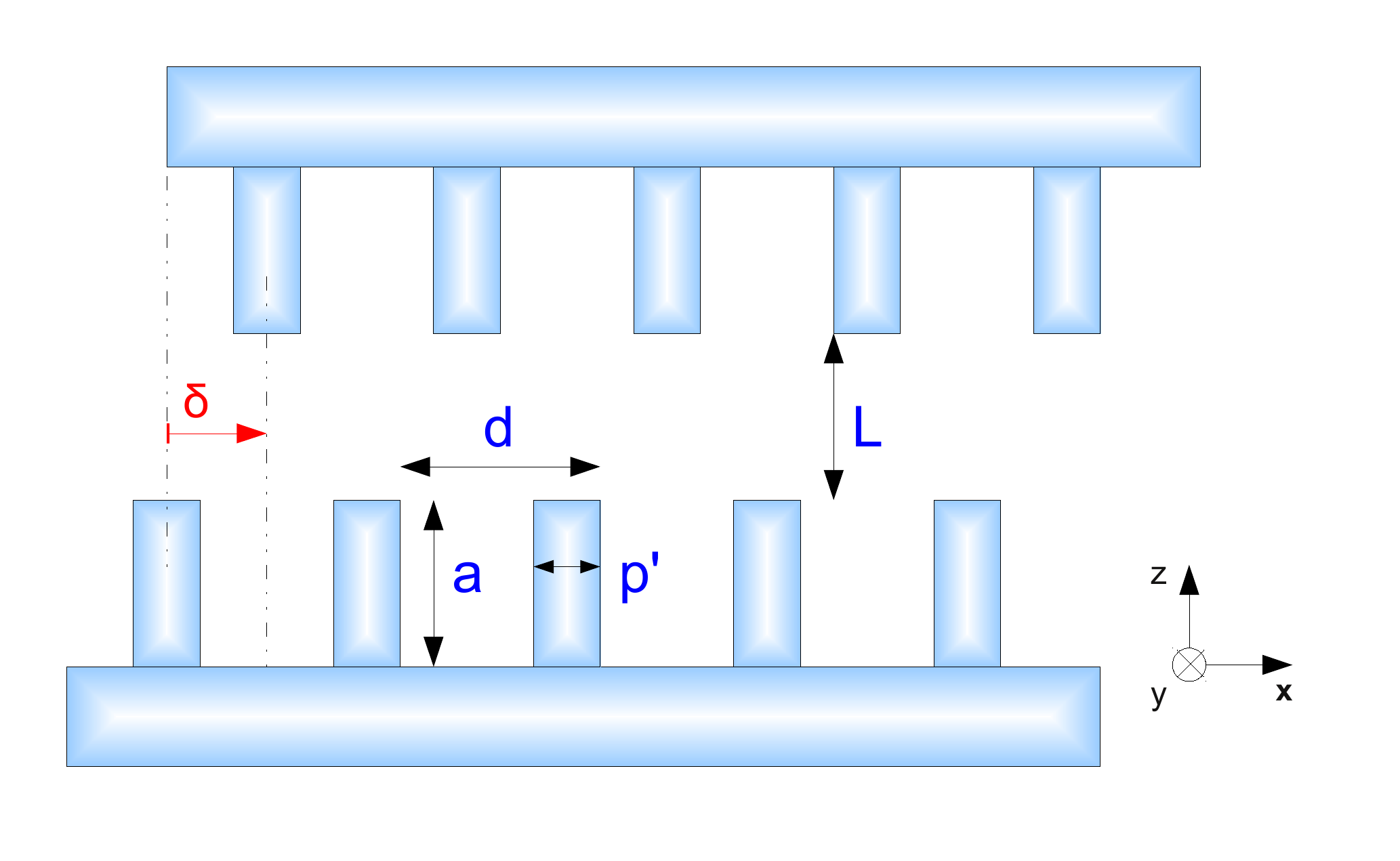}
\caption{\label{fig:epsart1} Two identical gratings facing each other at a distance $L$ and relatively shifted by a lateral displacement $\delta$.
The corrugations have a period $d$, height $a$ and thickness $p'$. The filling factor $p=p'/d$ is given as a percentage of the period $d$.}
\end{figure}
This is all the most interesting as a simple argument based on the
Proximity Approximation suggests a strong modulation of the
flux. Indeed, by assuming that one can use locally the plane-plane
heat transfer coefficient, it is seen that the flux is maximum for
$\delta=0$. The validity of the proximity approximation has been
discussed in the context of a plane-sphere \cite{Otey2011} and
between two spheres \cite{Narayanaswamy2008,Sasihithlu2011}. While it appears to be
valid for radius of curvatures larger than $20\mu$m, this validity
in the context of lamellar gratings with sub-wavelength periods
remains an open question. Here, we investigate this issue by using
the exact formalism of scattering theory. Furthermore, we discuss
the physical phenomena involved and show that the nature of the 
material needs to be taken into account when
discussing the validity of the Proximity Approximation.

Based on the scattering formalism developed in~\cite{Bimonte2009},
we consider two corrugated profiles at temperatures $T_{1}$ and
$T_{2}$, as shown in Fig.1. The heat transfer is constructed from
the statistical average of the $(x,y)$ sum over the $z$ component of
the Poynting vector $S_{z}$ and is thus related to a flux. We define
the wave vector $\mathbf{k}=(\mathbf{k}_{\perp}, k_{z})$ with
$k_{z}=\sqrt{\omega^{2}/c^{2}-\mathbf{k}_{\perp}^{2}}$ defined with 
$-\pi/2 < \arg k_z \leqslant \pi/2 $.

Following~\cite{Lambrecht2008}, we then introduce the reflection
operators $\mathbf{R_{1}(\omega)}$ and $\mathbf{R_{2}(\omega)}$ of
the two gratings separated by a distance $L$, by which we understand
the distance of closest approach, equal to zero at contact. We then
set our scattering operators such that
$\mathbf{S_{1}}=\mathbf{R_{1}(\omega)}$ and $\mathbf{S_{2}} =
e^{\imath k_{z} L} \mathbf{R_{2}(\omega)} e^{\imath k_{z} L}$.
According to the scattering formalism for gratings developed
in~\cite{Lambrecht2008,Bao2010}, the scattering matrices are of
dimensions $2(2N+1)$, where $N$ is the order of diffraction.

We now define the operators $\mathbf{\mathbf{\Sigma}}_{n}^{p \omega / e \omega}=\frac{1}{2} k_{z}^n \Pi^{p \omega / e \omega}$ as constructed from the projectors
on the propagative and evanescent sectors, respectively :
\begin{eqnarray}
\Pi_{\alpha \alpha'}^{p \omega}=\delta_{\alpha \alpha'} [1 + \text{sgn}(\omega^{2}/c^{2}- \mathbf{k}_{\perp}^{2})]
\label{eq:e1}
\\
\Pi_{\alpha \alpha'}^{e \omega}=\delta_{\alpha \alpha'} [1 - \text{sgn}(\omega^{2}/c^{2}- \mathbf{k}_{\perp}^{2})]
\label{eq:e2}
\end{eqnarray}

\noindent where $\alpha=s,p$ represents the transverse electric and transverse magnetic polarizations, respectively. The thermal energy density per
field mode at temperature $T$ writes $e_{T}(\omega) = \hslash \omega / \left( e^{\hbar \omega / k_{B} T} -1 \right)$.
We can now express the heat transfer coefficient between two gratings of same corrugation depths $a$ as :
\begin{eqnarray}
h = \frac{1}{|T_{1}-T_{2}|} \int \frac{d\omega}{2\pi} \left( e_{T_1}(\omega) - e_{T_2}(\omega) \right) H_{12}
\label{eq:e3}
\end{eqnarray}

with
\begin{eqnarray}
& & H_{12} =\int_{k_{x}=-\pi /d}^{+ \pi /d}
\int_{k_{y} \in \mathbb{R}} \frac{dk_{x} dk_{y}}{4\pi^2} \text{tr}(\mathbf{D} \mathbf{W_{1}} \mathbf{D}^{\dag} \mathbf{W_{2}})
\label{eq:e4}\\ 
& & \mathbf{D} = (1-\mathbf{S_{1}} \mathbf{S_{2}})^{-1}
\label{eq:e5}\\
& & \mathbf{W_{1}} = \mathbf{\mathbf{\Sigma}}_{-1}^{p \omega} - \mathbf{S_{1}} \mathbf{\mathbf{\Sigma}}_{-1}^{p \omega} \mathbf{S_{1}}^{\dag} +
\mathbf{S_{1}} \mathbf{\mathbf{\Sigma}}_{-1}^{e \omega} - \mathbf{\mathbf{\Sigma}}_{-1}^{e \omega} \mathbf{S_{1}}^{\dag}
\label{eq:e6}\\
& & \mathbf{W_{2}} =  \mathbf{\mathbf{\Sigma}}_{1}^{p \omega} - \mathbf{S_{2}}^{\dag} \mathbf{\mathbf{\Sigma}}_{1}^{p \omega} \mathbf{S_{2}} +
\mathbf{S_{2}}^{\dag} \mathbf{\mathbf{\Sigma}}_{1}^{e \omega} - \mathbf{\mathbf{\Sigma}}_{1}^{e \omega} \mathbf{S_{2}}
\label{eq:e7}
\label{eq:e8}
\end{eqnarray}

It is noteworthy that the heat transfer depends on the shape and material properties of the gratings only through their scattering
matrices $\mathbf{S_{1}}$ and $\mathbf{S_{2}}$. Furthermore, the factor $e_{T_1}(\omega) - e_{T_2}(\omega)$ introduces a cut-off
for all frequencies larger than $k_B T/\hslash$. It is hence $H_{12}$ in equation (\ref{eq:e4}), which corresponds to the sum of
the transmission factors of the modes, that gives rise to the interesting modes pertaining to the near-field contribution.

Note also that the first perpendicular wave vector component $k_{x}$ belongs to the first Brillouin zone between $-\pi /d$ and $+ \pi /d$,
whereas $k_{y} \in \mathbb{R}$ is not restricted. A practical challenge of the numerical integration of $h$ lays in the choice of the
boundaries of $\omega$ and $k_{y}$ through a careful study of the integrand of equation (\ref{eq:e4}) plotted over the whole range of frequencies
to determine the modes.

\begin{figure}
\includegraphics[scale=0.7]{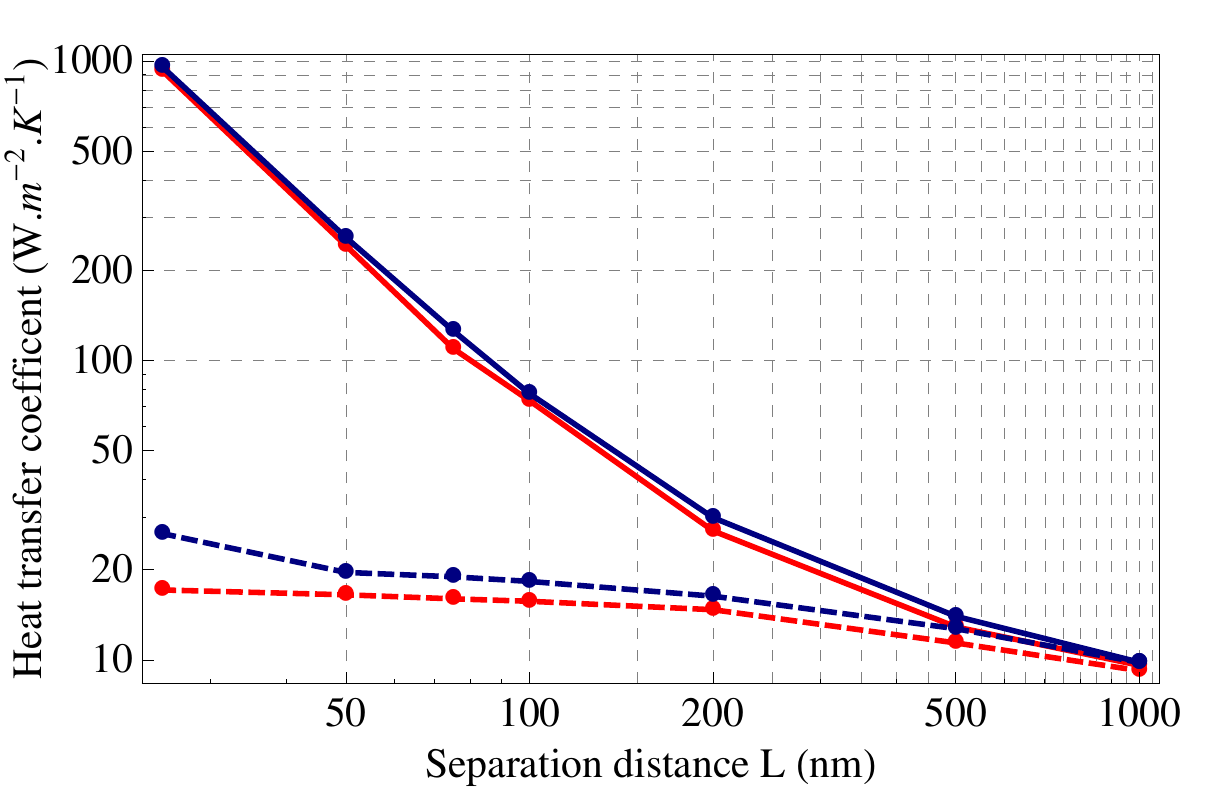}
\caption{\label{fig:epsart2} Heat transfer coefficients as a
function of separation distance $L$, when the gratings are not
laterally displaced (blue solid line) and when they are by half a
period (blue dashed line). This is compared with the Proximity
Approximation in red. The gratings have a period $d=1500$nm, filling
factor $p=20\%$, and groove depth $a=500$nm.}
\end{figure}

\begin{figure}
\includegraphics[scale=0.252]{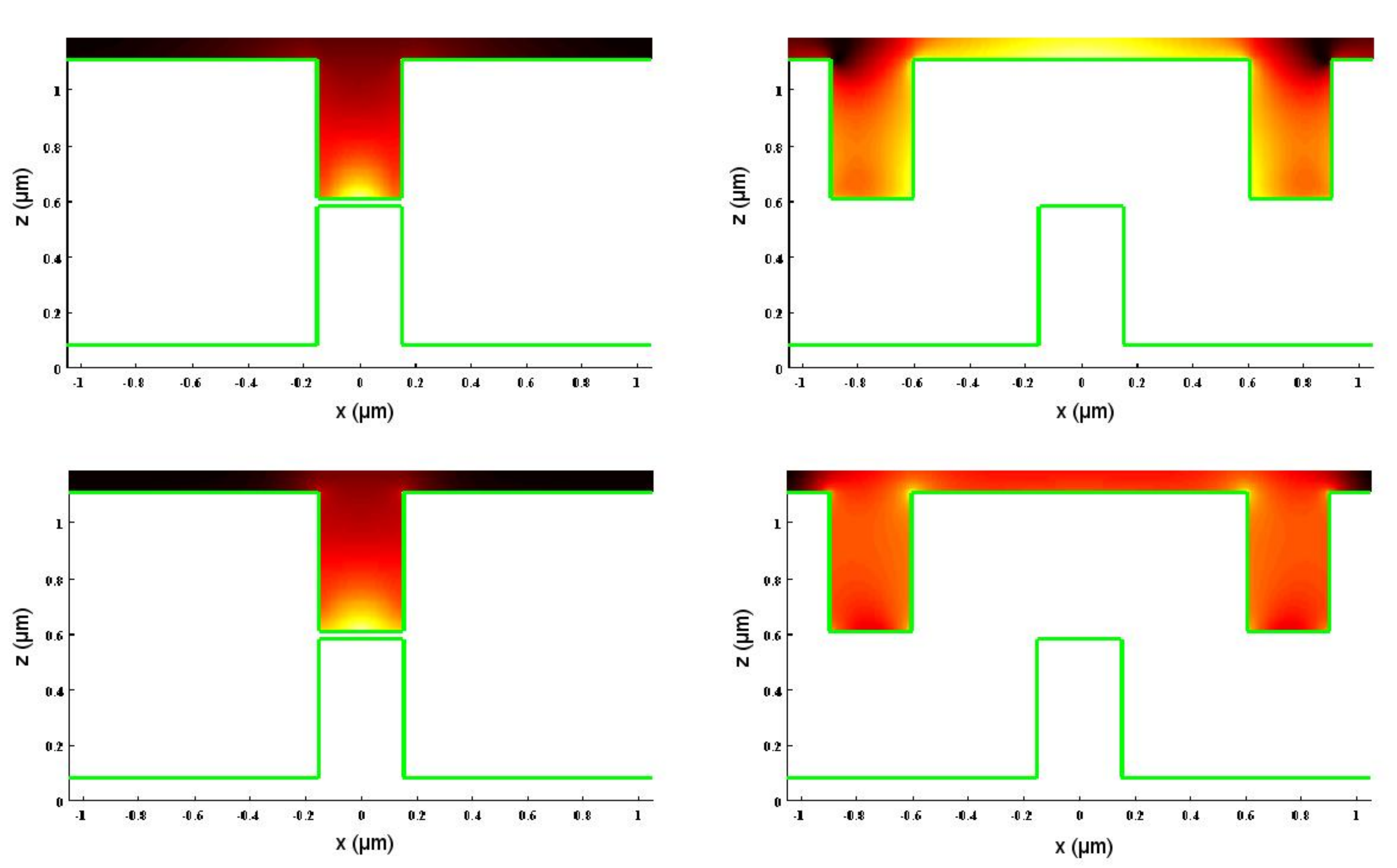}
\caption{\label{fig:epsart2b} Field modulus map of a given source
dipole placed in the middle of a corrugation and right under the
surface. The field is here represented only in the upper grating, 
so as to highlight where the absorption takes place. This is for 
gratings at a separation distance
$L=25$nm. The two figures on the left display the profiles in the
$xz-$plane (in green) when they are aligned ($\delta=0$), and the
two figures on the right when they are laterally displaced by
half-a-period ($\delta=d/2$), both for two different wavelengths
$\lambda=8.75 \mu$m (top) and $9.15 \mu$m (down).}
\end{figure}

We will from now on consider two gratings of silica glass SiO$_{2}$, 
the dielectric properties of
which are given in~\cite{PalikBOOK}. This material is chosen as it
supports surface phonon-polaritons, which are known to enhance the
flux. The gratings temperatures are supposed to be $T_{1}=310$ K and
$T_{2}=290$ K. Two sets of data are systematically computed : the
first corresponds to zero lateral displacement of the two plates
along the $x$-axis ($\delta=0$) so that the corrugation maxima
directly face those from the opposite profile. The second
corresponds to a lateral displacement of half the grating period
($\delta=d/2$), so that the corrugations maxima face the corrugation
trenches of the opposite profile. In near-field, the two structured
plates expose a larger surface to each other at $\delta=0$ than at
$\delta=d/2$, so that we expect a strong modulation of the heat
transfer coefficient which will be discussed later. This is based on
the assumption that the plane-plane heat transfer coefficient is
locally valid.

The results of the scattering approach can be compared with the PA,
which consists on the weighted sum of the planar normal
contributions $h_0(L)$ depending on the local separation
distances $L$ within each period. Assuming that $p<50\%$,
we have for $\delta \leq p'$ :

\begin{eqnarray}
h^{\text{PA}}_{\delta}(L) = & & \frac{p'-\delta}{d} h_0(L) + \frac{2\delta}{d} h_0(L+a) \\
& & + \left(1-\frac{p'+\delta}{d}\right) h_0(L+2a) \notag
\label{eq:e9}
\end{eqnarray}
For $\delta > p$, we find the following saturation value of :

\begin{eqnarray}
h^{\text{PA}}_{p'}(L) = & & \frac{2p'}{d} h_0(L+a) \\
& & + \left(1-\frac{2p'}{d}\right) h_0(L+2a) \notag
\label{eq:e10}
\end{eqnarray}

In what follows, we study in detail the interplay between surface
waves and corrugations. In particular, we assess the validity of the
PA. Fig.2 shows the heat transfer coefficient for $\delta=0$ and
$\delta=d/2$, as a function of the separation distance $L$, for two
gratings of period $d=1500$nm, filling factor $p=20\%$, and groove
depth $a=500$nm. The results are compared with the PA. Regardless of
the distance, we can see that the PA is a good approximation to the
heat transfer coefficient at $\delta=0$, but not at $\delta=d/2$. At
$L=25$nm, the error of the PA is of $\sim3\%$ for $\delta=0$, and of
$\sim35\%$ for $\delta=d/2$.

The reason for this is illustrated on Fig.3, which shows the field
modulus map for a given source dipole that is placed in the middle
of a corrugation right under the surface, and which is oriented
perpendicular to it. The color scale is logarithmic. The intensity or 
square modulus of the electric field 
is represented only in the upper grating so as to highlight the place 
of absorption. The gratings
have a separation distance $L=25$nm, corrugation depth $a=500$nm,
period $d=1500$nm, and filling factor $p=20\%$. Two different
wavelengths $\lambda=8.75\mu$m and $\lambda=9.15\mu$m are
considered, knowing that SiO$_2$ has two resonance frequencies at
$\lambda=8.75\mu$m and $\lambda=21\mu$m. In the case where $\delta=0$ and
$\lambda=8.75\mu$m, we see that the field is clearly both intense
and confined. As $8.75\mu$m corresponds to the horizontal asymptote
of the surface phonon dispersion relation, a large number of modes
with different values of the wave vector are excited. 
This leads to a highly localized subwavelength hot spot. At 
$9.15 \mu$m, the spot is broader as expected : this is similar to the 
loss of resolution of surperlens away from the resonance. On the 
right column of the figure, we show the intensity for $\delta=d/2$. 
It is seen that the heated region is delocalized so that PA is 
clearly not valid. In this regime, the heat transfer is no longer 
due to a dipole-dipole interaction through the gap. Instead, a 
dipole excites modes of the structures. In turn, these spatially 
extended modes produce dissipation in the walls. This discussion 
indicates that PA is valid if the gap width does not vary 
significantly on a length scale given by the spatial extension of 
the modes.

\begin{figure}
\includegraphics[scale=0.71]{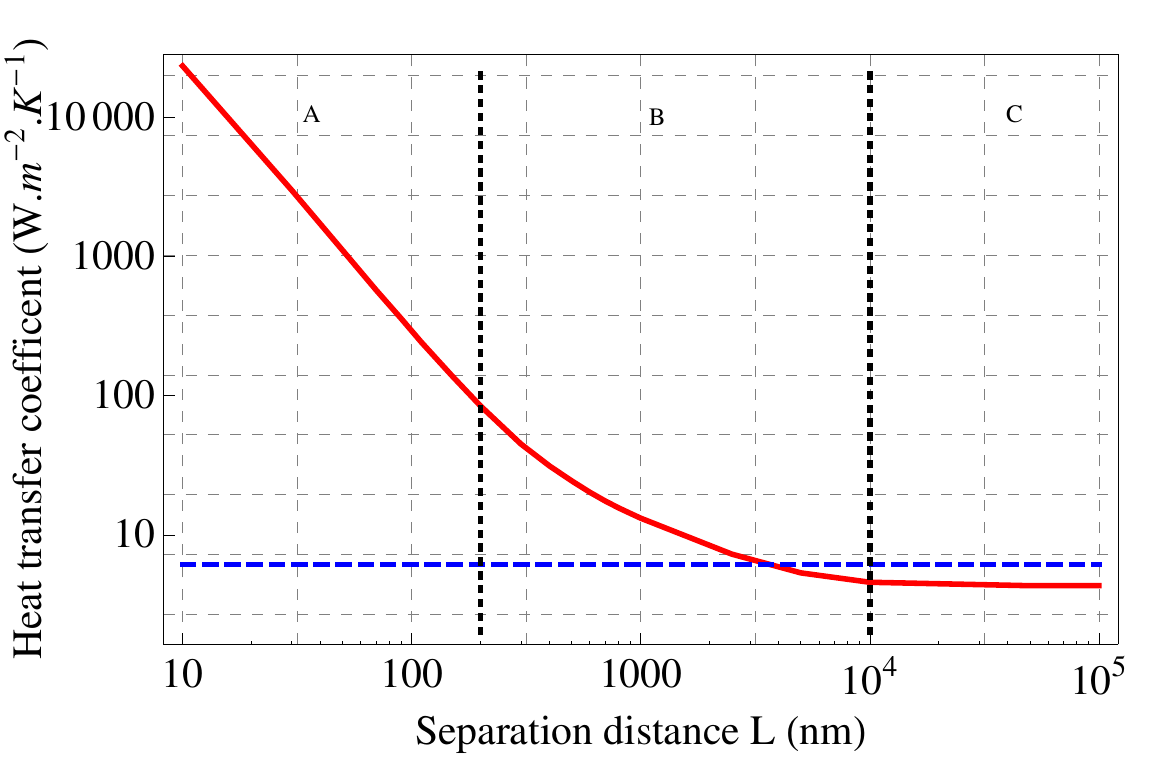}
\caption{\label{fig:epsart3} Heat transfer coefficients as a
function of the separation distance $L$ between two plane mirrors of
SiO$_2$ (red solid curve), compared with the black body limit (blue
dashed line). One can divide the separation distance in three
domains A, B, and C, respectively corresponding to the extreme
near-field below $200$nm, to the near-field from $200$nm to
$10\mu$m, and to the domain of Stefan-Boltzmann's law beyond
$10\mu$m. This can be seen by the change of the slope of the curve
along these three ranges.}
\end{figure}

To further illustrate this qualitative dependence of the radiative
heat transfer on separation distance, we show in Fig.4 the heat
transfer coefficients as a function of the separation distance $L$
between two plates of SiO$_2$. One can distinguish three domains
$A$, $B$, and $C$, corresponding respectively to the extreme
near-field below $200$nm, to the near-field from $200$nm to
$10\mu$m, and to the domain of Stefan–Boltzmann's law beyond
$10\mu$m. The heat transfer coefficient changes in slope along these
three ranges: the strongest contributions come respectively from the
dipole-dipole interaction, from surface phonon-polaritons, and from
the classical radiative heat transfer. The
contribution in the first domain corresponds to the localized heat
transfer seen in the upper-left map of Fig.3, whereas the main
contribution in the second domain corresponds to the delocalized
heat transfer mediated by the surface wave seen on the right maps of
Fig.3.

\begin{figure}
\includegraphics[scale=0.72]{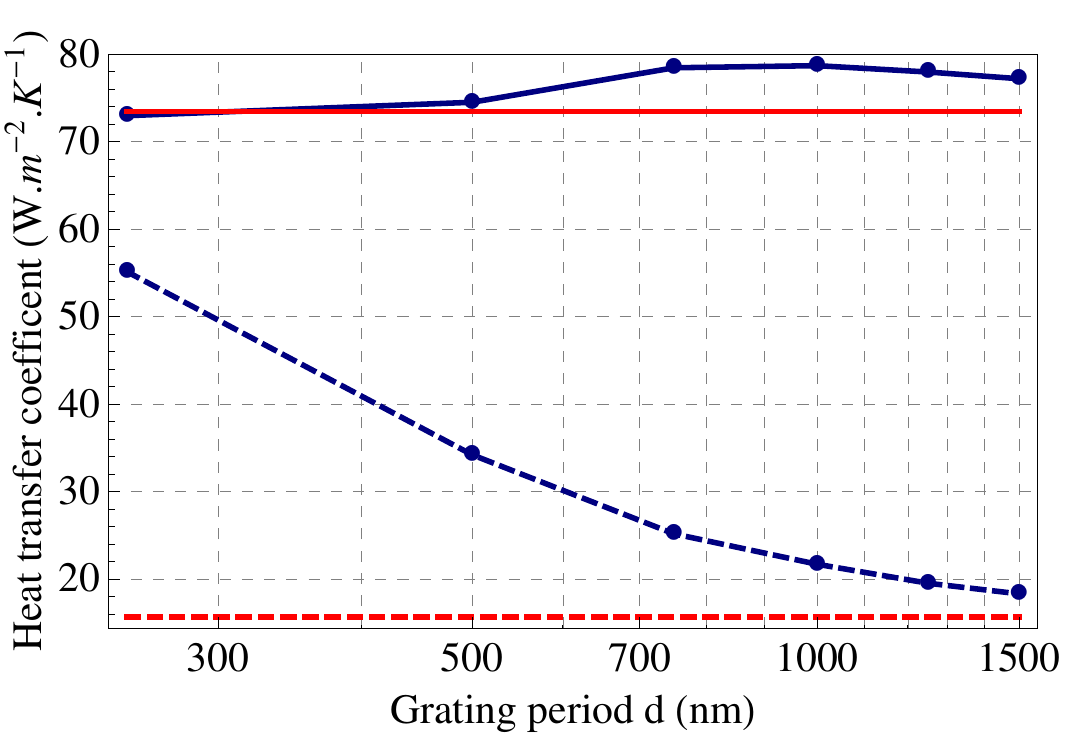}
\caption{\label{fig:epsart4} Heat transfer coefficients as a
function of grating period $d$, when the gratings are not laterally
displaced (solid blue line) and when they are displaced by half a
period (dashed blue line). This is compared with the PA in red. The
gratings have a groove depth $a=500$nm, filling factor $p=20\%$, and
are at a separation distance $L=100$nm.}
\end{figure}

It is also instructive to study the heat transfer modulation as a
function of the corrugation period $d$, as shown in Fig.5. We have
selected six types of gratings with corrugation periods ranging from
$d = 250$ to $1500$nm, each with a groove depth $a=500$nm and
filling factor still fixed at $p=20\%$. The separation distance is
$L=100$nm. The fact that the heat transfer coefficients at
$\delta=0$ does not vary much with a change of period is a further
confirmation of the validity of the PA in this configuration. At
$\delta=d/2$, however the scattering and PA results radically differ
for small periods, but tend to agree for large periods. The reason
for this is that when $d \rightarrow\infty$, the ratio $a/d$ tends
to zero and we expect the heat transfer to be well approximated by
the plane-plane case, and hence the PA.
\begin{figure}
\includegraphics[scale=0.6]{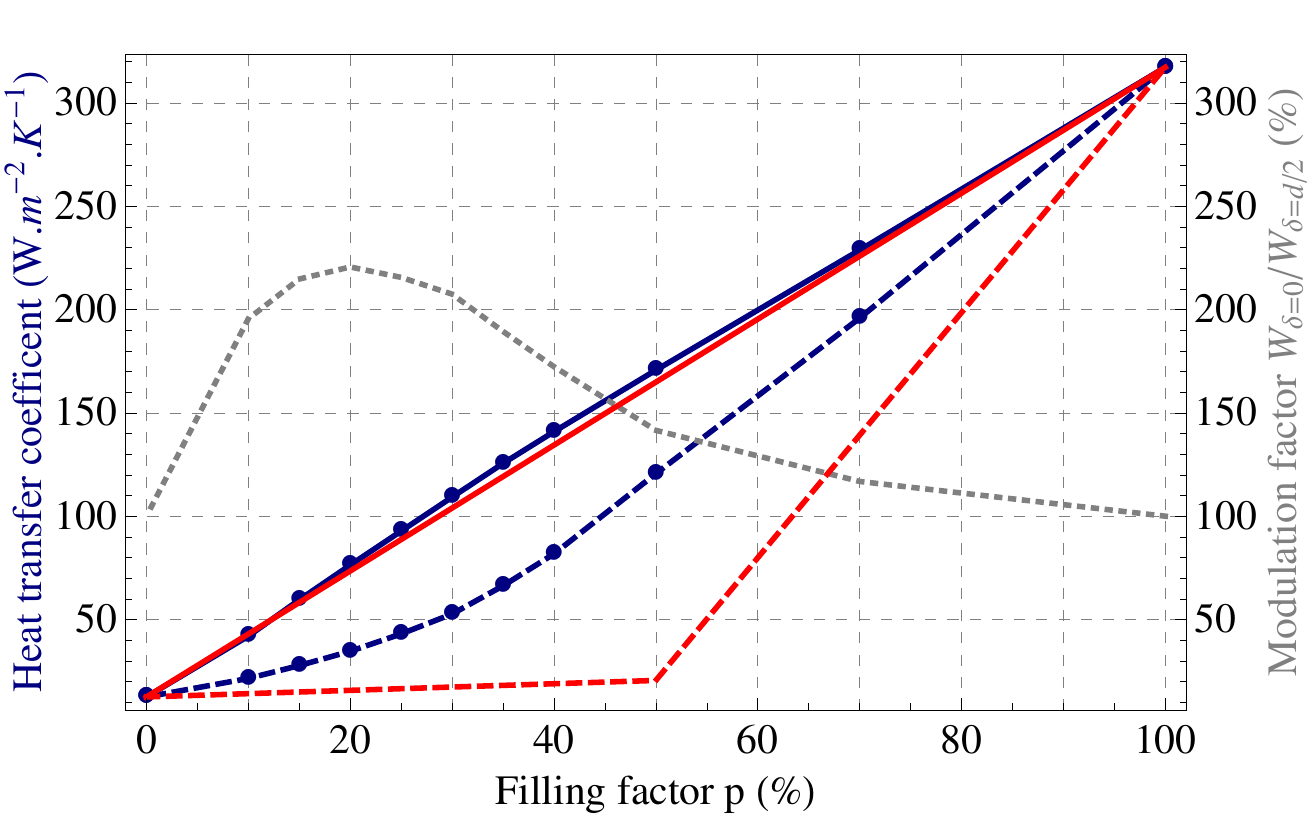}
\caption{\label{fig:epsart5} Heat transfer coefficients as a
function of filling factor $p$, when the gratings are not laterally
displaced (solid blue line) and when they are displaced by half a
period (dashed blue line). Respective PA predictions are in red. The
dotted gray line is the percentage of the modulation factor
$h_{\delta = 0} / h_{\delta = d/2}$. Gratings have a period and
groove depth of $500$nm, and are separated by a distance $L=100$
nm.}
\end{figure}

Let us finally turn to the discussion of the modulation effect.
Fig.2 shows that the heat transfer depends dramatically on the
lateral displacement of the two surfaces, opening the possibility of
a strong modulation via only lateral displacement of one of the two
plates at a fixed distance. To assess the possible performance of
such a system as a thermal modulator~\ref{Biehs2011}, we investigate the modulation
factor $h_{\delta = 0} / h_{\delta = d/2}$ for different filling
factors. The results are illustrated in Fig.6 for gratings with a
period and groove depth $a=500$nm, and a separation distance $L=100$
nm. For these large separations, the modulation factor $h_{\delta =
0} / h_{\delta = d/2}$ still reaches a maximum of about $2.2$, at a
filling factor corresponding to $20\%$ of the total grating period.
At short distances ($L\sim 25$nm) it can reach up to 35 (c.f.
Fig.2.).

So based on a - to our knowledge first - study of radiative heat transfer for
corrugated dielectric plates - we have clarified the origin of the
success and failure of the PA by analyzing the interplay between
surface waves resonances and corrugations. we have shown for various nanograting geometries and
separation distances that Derjaguin's Proximity Approximation is
clearly valid for $\delta=0$ but fails for $\delta=d/2$. The key to the
understanding of the system is the comparison of the lateral length
scale of the surface corrugation with the lateral extension of
surface waves involved in the heat transfer. Finally, we have
narrowed down the optimum geometrical parameters of a thermal
modulator device for nanosystems based on a lateral displacement of
two corrugated plates facing each other at fixed distance. We found
in general a stronger modulation for small filling factors and
separation distances, and for large grating periods. In certain
regimes it is possible to reach a modulation factor of more than
$35$. An in-depth study of the modes accounting for the most
important part of the heat transfer would be an interesting prospect
as well as to further enhance the modulation by using a broader
range of materials~\cite{VanZwolB2011} such as different alloys
combining the polaritons of certain dielectrics and the near-field
properties of metals. The issue of heat transfer in near-field in
the case of coatings~\cite{Biehs2007}, phase change
materials~\cite{VanZwol2011,VanZwolB2011},
metamaterials~\cite{Basu2009,Francoeur2011}, or graphene-covered
dielectrics~\cite{Svetovoy2012} in this regard should also be
explored.

The authors thank the ESF Research Networking Programme CASIMIR (www.casimir-network.com) for providing excellent possibilities for discussions
and exchange. The research described here has been supported by Triangle de la Physique contract EIEM 2010-037T.

\bibliography{References}

\end{document}